\input{aipcheck}

\documentclass[
    ,final            % use final for the camera ready runs
  ]
  {aipproc}

\usepackage{epstopdf}

\layoutstyle{8x11single}

\newcommand{\ket}{\,\rangle}
\newcommand{\bra}{\langle \,}
\newcommand{\Int}{\displaystyle{\int}}
\newcommand{\Frac}[2]{\frac{\displaystyle #1}{\displaystyle #2}}
\newcommand{\be}{\begin{equation}}
\newcommand{\ee}{\end{equation}}
\newcommand{\cO}{{\cal O}}

\newcommand{\nn}{\nonumber}

\begin{document}

\title{A few words about Resonances in the  \\ Electroweak Effective Lagrangian\footnote{Talk given at {\it XIth Quark Confinement and Hadron Spectrum}, 8-12th September (2014), Saint Petersburg (Russia). Preprint numbers: IFIC-15-04, IFT-UAM-CSIC-15-009, FTUAM-15-3, FTUV-15-2801.}}

\classification{12.39.Fe, 12.60.Fr, 12.60.Nz, 12.60.Rc} % Chiral Lagrangians, Extensions of electroweak Higgs sector, Technicolor models, Composite models of unification
\keywords      {Higgs Physics, Beyond Standard Model, Chiral Lagrangians, Technicolor, Composite Models}

\author{Ignasi Rosell}{
 address={Speaker \\ \phantom{hi}}
  ,altaddress={Departamento de Ciencias F\'\i sicas, Matem\'aticas y de la Computaci\'on, Universidad CEU Cardenal Herrera, \\ c/ Sant Bartomeu 55, 46115 Alfara del Patriarca, Val\`encia, Spain \\ \phantom{hi} }
}

\author{Antonio Pich}{
  address={Departament de F\'\i sica Te\`orica, IFIC, Universitat de Val\`encia -- CSIC, \\ Apt. Correus 22085, 46071 Val\`encia, Spain \\ \phantom{hi} }
}

\author{Joaqu\'\i n Santos}{
  address={Departament de F\'\i sica Te\`orica, IFIC, Universitat de Val\`encia -- CSIC, \\ Apt. Correus 22085, 46071 Val\`encia, Spain \\ \phantom{hi} }  
}

\author{Juan Jos\'e Sanz-Cillero}{
  address={Departamento de F\'\i sica Te\'orica and Instituto F\'\i sica Te\'orica, IFT-UAM/CSIC, \\ Universidad Aut\'onoma de Madrid, Cantoblanco, 28049 Madrid, Spain}
}

\begin{abstract}
Contrary to a widely spread believe, we have demonstrated that strongly coupled electroweak models including both a light Higgs-like boson and massive spin-$1$ resonances are not in conflict with experimental constraints on the oblique $S$ and $T$ parameters. We use an effective Lagrangian implementing the chiral symmetry breaking $SU(2)_L\otimes SU(2)_R\rightarrow SU(2)_{L+R}$ that contains the Standard Model gauge bosons coupled to the electroweak Goldstones, one Higgs-like scalar state $h$ with mass $m_{h}=126\,$GeV and the lightest vector and axial-vector resonance multiplets $V$ and $A$. We have considered the one-loop calculation of $S$ and $T$ in order to study the viability of these strongly-coupled scenarios, being short-distance constraints and dispersive relations the main ingredients of the calculation. Once we have constrained the resonance parameters, we do a first approach to the determination of the low energy constants of the electroweak effective theory at low energies (without resonances). We show this determination in the case of the purely Higgsless bosonic Lagrangian.

\end{abstract}

\maketitle

\section{Introduction}

A Higgs-like boson around $126\,$GeV was discovered at the LHC. Although its properties are being measured yet, it complies with the expected behaviour and therefore it is a very compelling candidate to be the Standard Model (SM) Higgs. An obvious question to address is to which extent alternative scenarios of Electroweak Symmetry Breaking (EWSB) can be already discarded or strongly constrained. In particular, what are the implications for strongly-coupled models where the electroweak symmetry is broken dynamically?

The existing phenomenological tests have confirmed the $SU(2)_L\otimes SU(2)_R\rightarrow SU(2)_{L+R}$ pattern of symmetry breaking, giving rise to three Goldstone bosons $\pi$ which, in the unitary gauge, become the longitudinal polarizations of the gauge bosons. When the $U(1)_Y$ coupling $g'$ is neglected, the electroweak Goldstone dynamics is described at low energies by the same Lagrangian as the QCD pions, replacing the pion decay constant by the EWSB scale $v=(\sqrt{2}G_F)^{-1/2} = 246\,$GeV~\cite{AB:80,longhitano}. In most strongly-coupled scenarios the symmetry is nonlinearly realized and one expects the appearance of massive resonances generated by the non-perturbative interaction.

The dynamics of Goldstones and massive resonance states can be analyzed in a generic way by using an effective Lagrangian, based on symmetry considerations. The theoretical framework is completely analogous to the Resonance Chiral Theory description of QCD at GeV energies~\cite{RChT}.

Using these techniques, we have investigated in Ref.~\cite{paper2}, and as an update of Ref.~\cite{paper},  the oblique $S$ and $T$ parameters~\cite{Peskin:92}, characterizing the new physics contributions in the electroweak boson self-energies, within strongly-coupled models that incorporate a light Higgs-like boson. Adopting a dispersive approach and imposing a proper high-energy behaviour, it has been shown there that it is possible to calculate $S$ and $T$ at the next-to-leading order, {\it i.e.} at one-loop. Note that these results do not depend on unnecessary ultraviolet cut-offs. We concluded that there is room for these models, but they are stringently constrained. The vector and axial-vector states should be heavy enough (with masses above the TeV scale), the mass splitting between them is highly preferred to be small and the Higgs-like scalar should have a $WW$ coupling close to the Standard Model one. Previous one-loop analyses can be found in Ref.~\cite{other}.

As a continuation~\cite{paper3}, and as a first approach to the determination of the low energy constants of the electroweak effective theory at low energies (without resonances), we do this estimation in the case of the purely Higgsless bosonic Lagrangian (withouth resonances), {\it i.e.} the Longhitano's Lagrangian~\cite{longhitano}.

\section{Constraining the Resonance Theory from Phenomenology} 

We have considered a low-energy effective theory containing the SM gauge bosons coupled to the electroweak Goldstones, one light scalar state $h$ with mass $m_{h} = 126$~GeV and the lightest vector and axial-vector resonance multiplets $V_{\mu\nu}$ and $A_{\mu\nu}$. We have only assumed the SM pattern of EWSB, {\it i.e.} the theory is symmetric under $SU(2)_L\otimes SU(2)_R$ and becomes spontaneously broken to the diagonal subgroup $SU(2)_{L+R}$. $h$ is taken to be singlet under $SU(2)_{L+R}$, while $V_{\mu\nu}$ and $A_{\mu\nu}$ are triplets. The underlying theory is also assumed to preserve parity in this analysis. To build the Lagrangian we have only considered operators with the lowest number of derivatives, as higher-derivative terms are either proportional to the equations of motion or tend to violate the expected short-distance behaviour~\cite{paper2}. In order to determine the oblique $S$ and $T$ parameters one only needs the interactions~\cite{paper2}
\begin{equation}\label{eq:Lagrangian}
\mathcal{L} \,=\,
\frac{v^2}{4}\bra  u_\mu  u^\mu \ket \left( 1 + \frac{2\kappa_W}{v} h\right)
+ \frac{F_V}{2\sqrt{2}} \bra V_{\mu\nu} f^{\mu\nu}_+ \ket
+ \frac{i G_V}{2\sqrt{2}} \bra  V_{\mu\nu} [u^\mu , u^\nu] \ket \nonumber 
+ \frac{F_A}{2\sqrt{2}} \bra A_{\mu\nu} f^{\mu\nu}_- \ket 
+ \sqrt{2} \lambda_1^{hA}  \partial_\mu h  \bra  A^{\mu \nu} u_\nu \ket \,, \phantom{\frac{1}{2}}
\end{equation}
plus the standard gauge boson and resonance kinetic terms. We have followed the notation of Ref.~\cite{paper2}. The first term in (\ref{eq:Lagrangian}) gives the Goldstone Lagrangian, present in the SM, plus the scalar-Goldstone interactions. For $\kappa_W=1$ one recovers the $h\to\pi\pi$ vertex of the SM. %Note that $\kappa_W$ is called $\omega,\kappa_Z$ or $a$ in other references. 

The oblique parameter $S$ receives tree-level contributions from vector and axial-vector exchanges \cite{Peskin:92}, while $T$ is identically zero at lowest-order (LO):
\begin{equation}
S_{\mathrm{LO}} = 4\pi \left( \frac{F_V^2}{M_V^2}\! -\! \frac{F_A^2}{M_A^2} \right)  \,,
\qquad\quad
T_{\mathrm{LO}}=0 \,.
\label{eq:LO}
\end{equation}

To compute next-to-leading order (NLO) contributions we have used the dispersive representation of $S$ introduced by Peskin and Takeuchi~\cite{Peskin:92}, whose convergence requires a vanishing spectral function at short distances:
\begin{equation}
S\, =\, \Frac{16 \pi}{g^2\tan\theta_W}\,
\Int_0^\infty \, \Frac{{\rm dt}}{t} \, [\, \rho_S(t)\, - \, \rho_S(t)^{\rm SM} \, ]\, , \label{Sintegral}
\end{equation}
with $\rho_S(t)\,\,$ the spectral function of the $W^3B$ correlator~\cite{paper2,paper,Peskin:92}.

The calculation of $T$  is simplified by noticing that, up to corrections of $\mathcal{O}(m_W^2/M_R^2)$, $T=Z^{(+)}/Z^{(0)}-1$, being $Z^{(+)}$ and $Z^{(0)}$ the wave-function renormalization constants of the charged and neutral Goldstone bosons computed in the Landau gauge~\cite{Barbieri:1992dq}. A further simplification occurs by setting $g$ to zero, which does not break the custodial symmetry, so only the $B$-boson exchange produces an effect in $T$. This approximation captures the lowest order contribution to $T$   in its expansion in powers of   $g$ and $g'$.

Requiring the $W^3 B$ spectral function $\rho_S(t)$ to vanish at high energies channel by channel leads to a good convergence of the Goldstone self-energies, at least for the cuts we have considered. Then, their difference obeys an unsubtracted dispersion relation, which enables us to compute $T$ through the dispersive integral~\cite{paper2},
\begin{eqnarray}
T &=& \Frac{4 \pi}{g'^2 \cos^2\theta_W}\, \Int_0^\infty \,\Frac{{\rm dt}}{t^2} \,
[\, \rho_T(t) \, -\, \rho_T(t)^{\rm SM} \,] \, , \label{Tintegral}
\end{eqnarray}
with $\rho_T(t)\,\,$ the spectral function of the difference of the neutral and charged Goldstone self-energies.

It is quite interesting to remark the main assumptions we have done in our approach:
\begin{enumerate}
\item Only operators with {\bf at most two derivatives} have been kept in the action. Considering the equations of motion, field redefinitions and the high-energy behavior of form factors, it is possible to justify the absence of higher derivative operators~\cite{paper2}. This procedure works very well in the QCD case~\cite{RChT}.
\item Only the {\bf lightest vector and axial-vector resonance} multiplets have been considered. %This is known to be a good approximation since contributions from higher states are suppressed by their masses. 
QCD phenomenology supports this ``single-resonance'' approximation, owing to the kinematical suppression of heavier resonance contributions~\cite{RChT}.
\item Only contributions to the dispersive relations of (\ref{Sintegral}) and (\ref{Tintegral}) coming from the {\bf lightest two-particle channels without heavy resonances} have been considered, {\it i.e.} two Goldstones or one Goldstone plus one Higgs-like scalar resonance for $S$ and the $B$ boson plus one Goldstone or one Higgs-like scalar resonance for $T$. Note that contributions from higher cuts are kinematically suppressed: the $1/t$ or $1/t^2$ weights in the sum rules (\ref{Sintegral}) and (\ref{Tintegral}) enhance the contribution from the lightest thresholds and suppress channels with heavy states~\cite{L10}. $V\pi$ and $A\pi$ contributions were shown to be suppressed in a previous Higgsless analysis~\cite{paper}. Again, it is known that this procedure gives a very good approximation to the corresponding integrals in the QCD case~\cite{L10}.
\item Unlike what happens in QCD, the underlying theory is not known. Therefore, although we have worked at lowest order in $g$ and $g'$, the perturbative {\bf chiral counting} in powers of momenta is not well defined. We only know that loops are suppressed ($\hbar$ counting in the loop expansion) and that it works in QCD in the framework of the $1/N_C$ expansion, with $N_C$ the number of colours.  
\end{enumerate}

\begin{figure}
\includegraphics[scale=0.28]{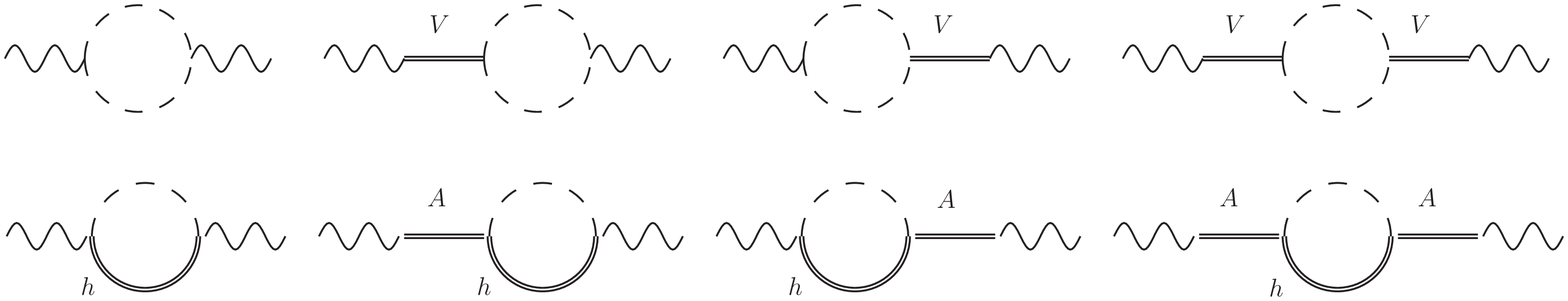}
\includegraphics[scale=0.28]{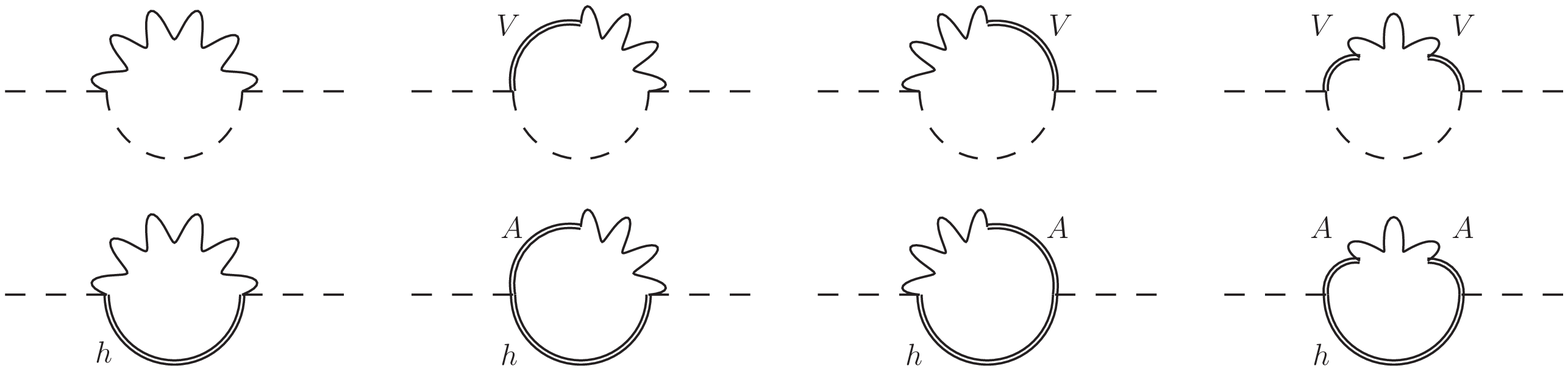}
\centering
\caption{NLO contributions to $S$ (left) and $T$ (right). A dashed (double) line stands for a Goldstone (heavy resonance or Higgs-like scalar) boson and a curved line represents a gauge boson.}
\label{fig-1}       
\end{figure}

Figure~\ref{fig-1} shows the computed one-loop absorptive contributions to $S$ and $T$.

\subsection{High-energy constraints}

Fixing $m_{h}=126$~GeV, one has seven undetermined parameters: $M_V$, $M_A$, $F_V$, $F_A$, $\sigma_V\equiv F_VG_V/v^2$, $\sigma_A\equiv F_A \lambda^{hA}_1/(\kappa_W v)$ and $\kappa_W$. This number can be reduced using short-distance information~\cite{paper2}:
\begin{enumerate}
\item {\bf  Vector form factor}. The two-Goldstone matrix element of the vector current defines the vector form factor. Imposing that it vanishes at $s\rightarrow \infty$, one finds~\cite{RChT}:
\begin{equation}
 \sigma_V \,\equiv\, \frac{F_V G_V}{v^2}\,=\,1\,. \label{VFF}
\end{equation}
\item {\bf Axial form factor}. The scalar-Goldstone matrix element of the axial-vector current defines the axial form factor. Imposing that it vanishes at $s\rightarrow \infty$, one finds~\cite{paper2,L10}:
\begin{equation}
\sigma_A \equiv \frac{F_A \lambda_1^{hA}}{\kappa_W\, v} \,=\,1\,. \label{AFF}
\end{equation}
\item {\bf Weinberg Sum Rules (WSRs)}. At leading-order the first and the second Weinberg sum rules~\cite{WSR} imply, respectively, 
\begin{equation}
F_{V}^{2} \, -\, F_{A}^{2}\,= \, v^2 \, , \qquad \qquad
F_{V}^{2}\, M_{V}^{2} \, -\, F_{A}^{2}\, M_{A}^{2} \,= \, 0   \, . \label{WSR}
\end{equation}
Finally, and once (\ref{VFF}) and (\ref{AFF}) have been considered, the second WSR implies at next-to-leading order
\begin{equation}
\kappa_W \,=\,  M_V^2/M_A^2 \,. \label{2WSRIm}
\end{equation}
Note that a small splitting between the vector and axial-vector resonances would imply $\kappa_W\sim 1$, that is, close to the SM value.
\end{enumerate}
As a conclusion, we have seven resonance parameters and up to five constraints. Taking into account that the second WSR is questionable in some scenarios, we have also studied the consequence of discarding the second WSR. 

\subsection{Phenomenology}

\begin{figure}
\centering
\includegraphics[scale=0.55]{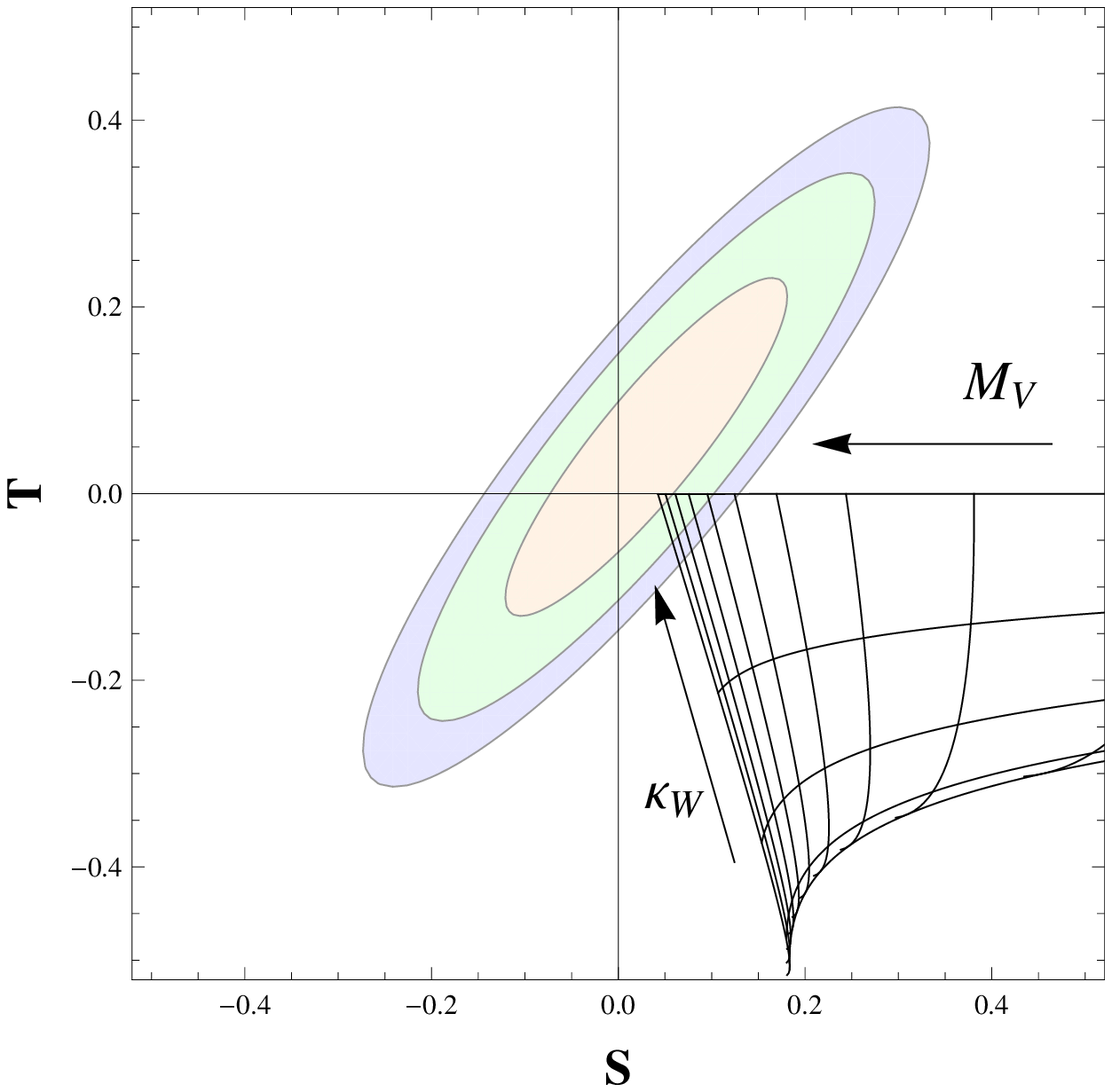} \quad \includegraphics[scale=0.60]{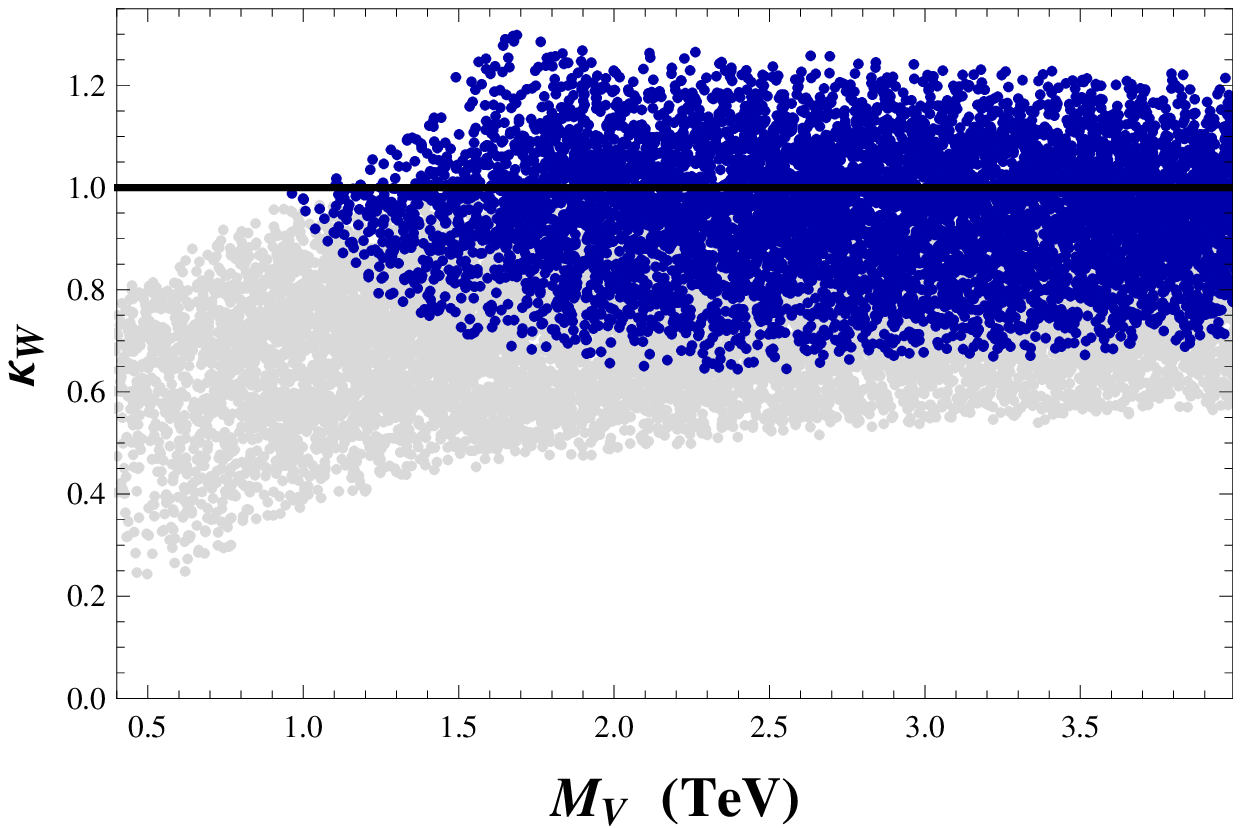}
\caption{{\bf NLO determinations of $S$ and $T$, imposing the two WSRs (left)}.
The approximately vertical curves correspond to constant values of $M_V$, from $1.5$ to $6.0$~TeV at intervals of $0.5$~TeV. The approximately horizontal curves have constant values of $\kappa_W$: $0.00, \, 0.25, 0.50, 0.75, 1.00$. The ellipses give the experimentally allowed regions at 68\%, 95\% and 99\% CL. {\bf Scatter plot for the 68\% CL region, in the case when only the first WSR is assumed (right)}. The dark blue and light gray regions correspond, respectively,  to $0.2<M_V/M_A<1$ and $0.02<M_V/M_A<0.2$. } \label{fig-2}       
\end{figure}

We have taken the SM reference point at $m_h = 126$ GeV, so the global fit gives the results $S = 0.03\pm 0.10$ and $T=0.05\pm0.12$, with a correlation coefficient of $0.891$~\cite{phenomenology}.
\begin{enumerate}
\item {\bf LO}. Considering the first and the second WSRs, $S_{\mathrm{LO}}$ becomes~\cite{Peskin:92}
\begin{equation}
S_{\mathrm{LO}} = \frac{4\pi v^2}{ M_V^{2}} \, \left( 1 + \frac{M_V^2}{M_A^2} \right) \,.
\end{equation}
Since the WSRs imply $M_A>M_V$, the prediction turns out to be bounded by  $4\pi v^2/M_V^{2} < S_{\rm   LO}  <  8 \pi v^2/M_V^2$~\cite{paper}. If only the first WSR is considered, and assuming  $M_A>M_V$, one obtains for $S$ the lower bound~\cite{paper}
\begin{equation}
S_{\mathrm{LO}} = 4\pi \left\{ \frac{v^2}{M_V^2}+ F_A^2 \left( \frac{1}{M_V^2} - \frac{1}{M_A^2} \right) \right\} > \frac{4\pi v^2}{M_V^2}. 
\end{equation}
The resonance masses need to be heavy enough to comply with the experimental bound, this is, much higher than the Higgs mass. From this point of view it is interesting to note that the Higss mass $m_{h} = 126\,$GeV is light in comparison with those resonances and the EW scale $\Lambda_{EW}= 4\pi v \sim 3\,$TeV. One finds a big gap between the lightest two particle cuts and the next ones (including vector and axial-vector resonances)~\cite{paper2}. As it has been explained previously, one expects therefore the NLO corrections to S to be widely dominated by the $\pi\pi$ and $h\pi$ cuts.
\item {\bf NLO with the 1st and the 2nd WSRs.} With (\ref{VFF})-(\ref{2WSRIm}) five of the seven resonance parameters are fixed and $S$ and $T$ are given in terms of $M_V$ and $M_A$ (or $M_V$ and $\kappa_W$)~\cite{paper2}:
\begin{eqnarray}
S & = & 4 \pi v^2 \left(\frac{1}{M_{V}^2}+\frac{1}{M_{A}^2}\right) + \frac{1}{12\pi} \bigg[ \log\frac{M_V^2}{m_{h}^2}  -\frac{11}{6} %\nonumber \\ &
+\frac{M_V^2}{M_A^2}\log\frac{M_A^2}{M_V^2}  - \frac{M_V^4}{M_A^4}\, \bigg(\log\frac{M_A^2}{m_{h}^2}-\frac{11}{6}\bigg) \bigg] \,,\quad \nonumber \\
T&= &    \frac{3}{16\pi \cos^2 \theta_W} \bigg[ 1 \!+\! \log \frac{m_{h}^2}{M_V^2}  \!-\! \frac{M_V^2}{M_A^2} \!\left( 1 \!+\! \log \frac{m_{h}^2}{M_A^2} \right)\! \bigg]  ,
\label{eq:T}
\end{eqnarray}
where terms of $\mathcal{O}(m_{h}^2/M_{V,A}^2)$ have been neglected. 

In Figure~\ref{fig-2} (left) we show the compatibility between the ``experimental'' values and these determinations~\cite{paper2}. The Higgs-like scalar should have a $WW$ coupling very close to the SM one. At 68\% (95\%) CL, one gets
$\kappa_W\in [0.97,1]$  ($[0.94,1]$), in nice agreement with LHC evidence, but more restrictive. Moreover, the vector and axial-vector states should be very heavy (and quite degenerate); one finds $M_V> 5$~TeV ($4$~TeV) at 68\% (95\%) CL.
\item {\bf NLO with the 1st WSR.} With (\ref{VFF}), (\ref{AFF}) and the first equation of (\ref{WSR}) one can still determine $T$ and obtain a lower bound of $S$ in terms of $M_V$, $M_A$ and $\kappa_W$~\cite{paper2}:
\begin{eqnarray}
S & \geq &    \frac{4 \pi v^2}{M_{V}^2} + \frac{1}{12\pi}  \bigg[ \log\frac{M_V^2}{m_{h}^2} -\frac{11}{6} %\nonumber \\ &\qquad \qquad 
  - \kappa_W^2 \bigg(\log\frac{M_A^2}{m_{h}^2}-\frac{17}{6} + \frac{M_A^2}{M_V^2}\!\bigg) \bigg]  , \nonumber \\
 T&=&    \frac{3}{16\pi \cos^2 \theta_W} \bigg[ 1 \!+\! \log \frac{m_{h}^2}{M_V^2}  \!-\! \kappa_W^2 \! \left( 1 \!+\! \log \frac{m_{h}^2}{M_A^2} \right)\!  \bigg]   ,
\label{eq:Tbis}
\end{eqnarray}
where $M_V<M_A$ has been assumed and again terms of $\mathcal{O}(m_{h}^2/M_{V,A}^2)$ have been neglected.

Figure~\ref{fig-2} (right) gives the allowed 68\% CL region in the space of parameters $M_V$ and $\kappa_W$, varying $M_V/M_A$ between $0.02$ and $1$~\cite{paper2}. Note, however, that values of $\kappa_W$ very different from the SM can only be obtained with a large splitting of the vector and axial-vector masses. In general, there is no solution for $\kappa_W >1.3$. Requiring $0.5<M_V/M_A<1$, leads to $\kappa_W > 0.84$ at 68\% CL, while the allowed vector mass stays above $1.5$~TeV.
\end{enumerate}

To sum up, the principal conclusions of this analysis have been the following ones~\cite{paper2}:
\begin{enumerate}
\item Strongly-coupled electroweak models with massive resonance states are still allowed by the current experimental data. In any case, these models are stringently constrained.
\item The Higgs-like boson with mass $m_{h}=126$~GeV must have a $WW$ coupling close to the SM one ($\kappa_W=1$). In those scenarios, such as asymptotically-free theories, where the second WSR is satisfied, the $S$ and $T$ constraints force $\kappa_W$ to be in the range $\left [ 0.94, 1\right]$ at 95\% CL, as shown in Figure~\ref{fig-2} (left). From Figure~\ref{fig-2} (right) it follows that larger departures from the SM value can be accommodated when the second WSR does not apply, but one needs to introduce a correspondingly large mass splitting between the vector and axial-vector states.
\item The vector and axial-vector states should be heavy enough (above the TeV scale), see Figure~\ref{fig-2}.
\item The mass splitting between the vector and axial-vector resonance fields is very small when the second WSR is valid (consider (\ref{2WSRIm}) and the restrictions on $\kappa_W$). % In any case, if one does not use the second WSR, this splitting is preferred to be small. 
\end{enumerate}

\section{Constraining the Electroweak Effective Theory from the Resonance Theory}

Once we have constrained the Resonance Theory by using short-distance constraints and the phenomenology, we want to use the Resonance Theory to determine the Low Energy Constants (LECs) of the electroweak effective theory at low energies (without resonances)~\cite{paper3}. As we have pointed out above, and as a first approach to this issue, we do this estimation in the case of the purely Higgsless bosonic case (without fermions). This exercise is similar to the estimation of the LECs of Chiral Perturbation Theory by using Resonance Chiral Theory~\cite{RChT}.

At {\bf high energies} we consider the Lagrangian (\ref{eq:Lagrangian}), whereas at {\bf low energies} we need to consider $\cO(p^4)$ operators without resonances or fermions~\cite{longhitano}: 
\begin{eqnarray}
\Delta \mathcal{L}_4 & = &
   \Frac{1}{4} a_1 \bra {f}_+^{\mu\nu} {f}_{+\, \mu\nu} - {f}_-^{\mu\nu} {f}_{-\, \mu\nu}\ket 
+\Frac{i}{2} (a_2-a_3) \bra {f}_+^{\mu\nu} [u_\mu, u_\nu] \ket  %+ \Frac{i}{2} (a_2+a_3) \bra {f}_-^{\mu\nu} [u_\mu, u_\nu] \ket 
+  a_4 \bra u_\mu u_\nu\ket \, \bra u^\mu u^\nu\ket 
\nn\\ &&\qquad 
+  a_5 \bra u_\mu u^\mu\ket^2 
+ \Frac{1}{2} H_1\bra {f}_+^{\mu\nu} {f}_{+\, \mu\nu} + {f}_-^{\mu\nu} {f}_{-\, \mu\nu}\ket  %+ \widetilde{H}_1 \bra {f}_+^{\mu\nu} {f}_{-\, \mu\nu} \ket 
\, , \label{eq.L4-Longhitano}
\end{eqnarray}
where we use the same notation as before and assume that parity is a good symmetry of the bosonic sector.

%At {\bf high energies} we consider $\cO(p^2)$ operators with heavy scalar ($S_1$), vector ($V$) and axial-vector ($A$) resonances and without fermions:
%\begin{eqnarray}
%\Delta \mathcal{L}_4' &=&  \frac{c_{d}}{\sqrt{2}} S_1 \langle u_\mu u^\mu \rangle + \frac{F_V}{2\sqrt{2}} \langle V_{\mu\nu} f_+^{\mu\nu}\rangle  +
%\frac{i G_V}{2\sqrt{2}} \langle V_{\mu\nu} [u^\mu, u^\nu] \rangle + \frac{\widetilde{F}_V}{2\sqrt{2}} \langle V_{\mu\nu} f_-^{\mu\nu}\rangle \nn  \\ &&
%+ \frac{F_A}{2\sqrt{2}} \langle A_{\mu\nu} f_-^{\mu\nu}\rangle +  \frac{\widetilde{F}_A}{2\sqrt{2}} \langle A_{\mu\nu} f_+^{\mu\nu}\rangle +
%\frac{i\widetilde{G}_A}{2\sqrt{2}} \langle A_{\mu\nu} [u^\mu, u^\nu] \rangle  \,.
%\end{eqnarray}
%Note that in the electroweak case we have more terms than in the QCD case because %the Lagrangian has to be invariant only under CP (and not under C and P).
%now we have made our Lagrangian invariant under CP (and not under C and P separately), allowing the presence of P-odd operators (indicated with a tilde). Also %notice that $S_1$ denotes a heavy resonance in this section and must no be confused with the light Higgs-like scalar.  

Integrating out the heavy resonances in a similar way as Ref.~\cite{RChT} does in the QCD case, we get the result~\cite{paper3}:
 \begin{eqnarray}
a_1\,=\, - \frac{F_V^2}{4M_V^2} + \frac{F_A^2}{4M_A^2} \,,  \qquad 
%
%(a_2+a_3) \,=\, 0    \,, \qquad 
%
(a_2-a_3)\, =\, -  \frac{F_VG_V}{2M_V^2}     \,, \qquad 
%
%(a_4+a_5) \, =\, 0  \, , 
%\nn \\
%
a_4 \,=\, -a_5 \, =\,  \frac{G_V^2}{4M_V^2} \,, \qquad 
H_1\, =\, - \frac{F_V^2}{8M_V^2} - \frac{F_A^2}{8M_A^2}\, ,  
%
%\widetilde{H}_1 \,=\, 0.  \qquad \qquad  
 \label{matching}
\end{eqnarray}
% \begin{eqnarray}
%a_1= - \frac{F_V^2}{4M_V^2}+\frac{{\widetilde{F}_V}^2}{4M_V^2} + \frac{F_A^2}{4M_A^2}-\frac{{\widetilde{F}_A}^2}{4M_A^2} \,,  \quad 
%
%(a_2+a_3) =   - \frac{\widetilde{F}_VG_V}{2M_V^2} - \frac{F_A\widetilde{G}_A}{2M_A^2}    \,, \quad 
%
%(a_2-a_3) = -  \frac{F_VG_V}{2M_V^2} - \frac{\widetilde{F}_A\widetilde{G}_A}{2M_A^2}    \,, \nn \\
%
% a_4 = \frac{G_V^2}{4M_V^2} \!+\! \frac{{\widetilde{G}_A}^2}{4M_A^2}   , \quad 
%
%a_5 = \frac{c_{d}^2}{4M_{S_1}^2} \!-\!\frac{G_V^2}{4M_V^2} \!-\! \frac{{\widetilde{G}_A}^2}{4M_A^2}   , \quad 
%
%H_1 = \!-\! \frac{F_V^2}{8M_V^2}\!-\!\frac{{\widetilde{F}_V}^2}{8M_V^2} \!-\! \frac{F_A^2}{8M_A^2}\!-\!\frac{{\widetilde{F}_A}^2}{8M_A^2} , \quad 
%
%\widetilde{H}_1 = \!-\! \frac{F_V \widetilde{F}_V}{4M_V^2}\! -\! \frac{F_A \widetilde{F}_A}{4M_A^2} . 
%\end{eqnarray}
The use of short-distance constraints is again very important in order to reduce the number of resonance parameters. In this way, (\ref{VFF}) and (\ref{WSR}) allow to determine $F_V$, $F_A$ and $G_V$ in terms of $v$, $M_V$ and $M_A$, so (\ref{matching}) becomes:
\begin{eqnarray}
a_1\,=\, -\frac{v^2}{4} \left(\frac{1}{M_V^2} + \frac{1}{M_A^2} \right) \,,  \qquad 
%
%(a_2+a_3) \,=\, 0    \,, \qquad 
%
(a_2-a_3)\, =\, -  \frac{v^2}{2M_V^2}     \,, \qquad  \nn \\
%
%(a_4+a_5)\, =\, 0 \, ,  \nn \\
%
a_4\,=\, -a_5\, =\, \frac{v^2}{4} \left( \frac{1}{M_V^2} - \frac{1}{M_A^2} \right) \,, \qquad 
H_1\, =\, - \frac{v^2}{8 }\left( \frac{1}{M_V^2} - \frac{1}{M_A^2} + \frac{2}{M_A^2-M_V^2} \right) \, .  
%
%\widetilde{H}_1 \,=\, 0. 
 \label{matching2}
\end{eqnarray}

The next step is the consideration of operators with fermions and Higgs fields  (in progress~\cite{paper3}). In~\cite{paper3} we also study a more general effective Lagrangian invariant under CP (and not under C and P separately), allowing the presence of P-odd operators, not considered in (\ref{eq:Lagrangian}).

\begin{theacknowledgments}
 We wish to thank the organizers of the conference for the pleasant conference. This work has been supported in part
by the Spanish Government and the European Commission [FPA2010-17747, FPA2011-23778, FPA2013-44773-P, SEV-2012-0249 (Severo Ochoa Program), CSD2007-00042 (Consolider Project CPAN)], the Generalitat Valenciana [PrometeoII/2013/007] and the Comunidad de Madrid [HEPHACOS S2009/ESP-1473].
\end{theacknowledgments}


\begin{thebibliography}{9}

\bibitem{AB:80} 
T. Appelquist and C. Bernard, Phys.\ Rev.\ D {\bf 22} (1980) 200; \\
%
%\cite{Dobado:1990zh}
%\bibitem{Dobado:1990zh}
  A.~Dobado, D.~Espriu and M.~J.~Herrero,
  %``Chiral Lagrangians as a tool to probe the symmetry breaking sector of the SM at LEP,''
  Phys.\ Lett.\ B {\bf 255} (1991) 405.
  %%CITATION = PHLTA,B255,405;%%
  %149 citations counted in INSPIRE as of 03 Jul 2013

%\cite{Longhitano:1980iz}
\bibitem{longhitano}
  A.~C.~Longhitano,
  %``Heavy Higgs Bosons in the Weinberg-Salam Model,''
  Phys.\ Rev.\ D {\bf 22} (1980) 1166;
  %%CITATION = PHRVA,D22,1166;%%
%
%\cite{Longhitano:1980tm}
%\bibitem{Longhitano:1980tm}
%  A.~C.~Longhitano,
  %``Low-Energy Impact of a Heavy Higgs Boson Sector,''
  Nucl.\ Phys.\ B {\bf 188} (1981) 118.
  %%CITATION = NUPHA,B188,118;%%

\bibitem{RChT}
%\bibitem{RChTa}
  G.~Ecker, J.~Gasser, A.~Pich and E.~de Rafael,
%  {\it The role of resonances in chiral perturbation theory},
  Nucl.\ Phys.\ B {\bf 321} (1989) 311; \\
  %%CITATION = NUPHA,B321,311;%%
%\bibitem{RChTb}
  G.~Ecker {\it et al.}, % J.~Gasser, H.~Leutwyler, A.~Pich and E.~de Rafael,
  %{\it Chiral Lagrangians for massive spin 1 fields},
  Phys.\ Lett.\ B {\bf 223} (1989) 425; \\
  %%CITATION = PHLTA,B223,425;%%
%\bibitem{RChTc}
  V.~Cirigliano {\it et al.},
  % G.~Ecker, M.~Eidem\"uller, R.~Kaiser, A.~Pich and J.~Portol\'es,
  %{\it Towards a consistent estimate of the chiral low-energy constants},
  Nucl.\ Phys.\ B {\bf 753} (2006) 139.
% [hep-ph/0603205].
  %%CITATION = HEP-PH 0603205;%%

%\cite{Pich:2012dv}
\bibitem{paper2}
 A.~Pich, I.~Rosell and J.~J.~Sanz-Cillero,
  %``Viability of strongly-coupled scenarios with a light Higgs-like boson,''
Phys.\ Rev.\ Lett.\  {\bf 110} (2013) 181801;  \\
 % arXiv:1212.6769 [hep-ph].
  %%CITATION = ARXIV:1212.6769;%%
  %5 citations counted in INSPIRE as of 10 Jun 2013
 %\cite{Pich:2013fea}
%\bibitem{Pich:2013fea}
 % A.~Pich, I.~Rosell and J.~J.~Sanz-Cillero,
  %``Oblique S and T Constraints on Electroweak Strongly-Coupled Models with a Light Higgs,''
  JHEP {\bf 1401} (2014) 157.
 % [arXiv:1310.3121 [hep-ph]].
  %%CITATION = ARXIV:1310.3121;%%
  %9 citations counted in INSPIRE as of 13 Jan 2015  
  
%\cite{Pich:2012jv}
\bibitem{paper}
  A.~Pich, I.~Rosell and J.~J.~Sanz-Cillero,
  %``One-Loop Calculation of the Oblique S Parameter in Higgsless Electroweak Models,''
  JHEP {\bf 1208} (2012) 106.
%  [arXiv:1206.3454 [hep-ph]].
  %%CITATION = ARXIV:1206.3454;%%

\bibitem{Peskin:92}
  M. E. Peskin and T. Takeuchi,
  % {\it Estimation of oblique electroweak corrections},
  Phys.\ Rev.\ D {\bf 46} (1992) 381;
  Phys.\ Rev.\ Lett.\  {\bf 65} (1990) 964.

\bibitem{other}
%\cite{Matsuzaki:2006wn}
%\bibitem{Matsuzaki:2006wn}
  S.~Matsuzaki, R.~S.~Chivukula, E.~H.~Simmons and M.~Tanabashi,
  %``One-Loop Corrections to the S and T Parameters in a Three Site Higgsless Model,''
  Phys.\ Rev.\ D {\bf 75} (2007) 073002; \\
%  [hep-ph/0607191],
  %%CITATION = HEP-PH/0607191;%%
 %\bibitem{S-Isidori:08}
    R. Barbieri,  G. Isidori,  V.S. Rychkov and E. Trincherini,
  %  {\it Heavy Vectors in Higgs-less models},
    Phys.\ Rev.\ D {\bf 78} (2008) 036012; \\
%    [arXiv:0806.1624 [hep-ph]];\\
%
%\bibitem{S-Cata:10}
    O. Cata and J.F. Kamenik,
   % {\it ElectroWeak Precision Observables at One-Loop in Higgsless models},
    Phys.\ Rev.\ D {\bf 83} (2011) 053010; \\
%    [arXiv:1010.2226 [hep-ph]].
%
%\cite{Foadi:2012ga}
%\bibitem{Foadi:2012ga}
  R.~Foadi and F.~Sannino,
  %``S and T Parameters from a Light Nonstandard Higgs versus Near Conformal Dynamics,''
  Phys.\ Rev.\ D {\bf 87} (2013) 015008; \\
  %[arXiv:1207.1541 [hep-ph]].
  %%CITATION = ARXIV:1207.1541;%%
  %7 citations counted in INSPIRE as of 03 Jul 2013
%
%\bibitem{S-Orgogozo:11}
    A. Orgogozo and  S. Rychkov,
   % {\it Exploring T and S parameters in Vector Meson Dominance Models of
%    Strong Electroweak Symmetry Breaking},
JHEP {\bf 1203} (2012) 046;  %    [arXiv:1111.3534 [hep-ph]]; 
%
%\cite{Orgogozo:2012ct}
%\bibitem{Orgogozo:2012ct}
 % A.~Orgogozo and S.~Rychkov,
  %``The S parameter for a Light Composite Higgs: a Dispersion Relation Approach,''
  JHEP {\bf 1306} (2013) 014.
 % [arXiv:1211.5543 [hep-ph]].
  %%CITATION = ARXIV:1211.5543;%%
  %5 citations counted in INSPIRE as of 03 Jul 2013

\bibitem{paper3}
A.~Pich, I.~Rosell, Joaqu\'\i n Santos and J.~J.~Sanz-Cillero, work in progress.

%\cite{Barbieri:1992dq}
\bibitem{Barbieri:1992dq}
  R.~Barbieri {\it et al.},
  %``Two loop heavy top effects in the Standard Model,''
  Nucl.\ Phys.\ B {\bf 409} (1993) 105.
  %%CITATION = NUPHA,B409,105;%%

%\cite{Pich:2008jm}
%\bibitem{Pich:2008jm}
\bibitem{L10}
  A.~Pich, I.~Rosell and J.~J.~Sanz-Cillero,
  %``Form-factors and current correlators: Chiral couplings L(10)mu) **r(mu) and C(87)**r(mu) at NLO in 1/N(C),''
  JHEP {\bf 0807} (2008) 014.
 % [arXiv:0803.1567 [hep-ph]].
  %%CITATION = ARXIV:0803.1567;%%
  %40 citations counted in INSPIRE as of 27 Jun 2013

\bibitem{WSR}
 %\cite{Weinberg:1967kj}
%\bibitem{Weinberg:1967kj}
  S.~Weinberg,
  %``Precise relations between the spectra of vector and axial vector mesons,''
  Phys.\ Rev.\ Lett.\  {\bf 18} (1967) 507.
  %%CITATION = PRLTA,18,507;%%
  
 \bibitem{phenomenology}
GFITTER, A Generic Fitter Project for HEP Model Testing, http://gfitter.desy.de/; \\
LEP Electroweak Working Group, http://lepewwg.web.cern.ch/LEPEWWG/.
%



\end{thebibliography}
\end{document}